\documentclass[preprint,preprintnumbers,amsmath,amssymb,prb]{revtex4-1}
\usepackage{graphicx}
\usepackage{subfigure}
\newlength{\picwidth}
\begin{document}
\setlength{\picwidth}{0.8\linewidth}
\title{A second superconducting energy gap of Nb$_3$Sn observed by breakjunction point-contact spectroscopy}
\author{M.\ Marz}
\affiliation{Physikalisches Institut, Karlsruher Institut f\"ur Technologie, D-76131
Karlsruhe, Germany}
\author{W.\ Goldacker}
\affiliation{Institut f\"ur Technische Physik, Karlsruher Institut f\"ur Technologie, D-76131 Karlsruhe, Germany}
\author{R.\ Lortz}
\altaffiliation[present address: ]{Department of Physics, The Hong Kong University of Science \& Technology, Clear Water Bay, Kowloon, Hong Kong}
\affiliation{Department of Condensed Matter Physics, University of Geneva, CH-1211 Geneva 4, Switzerland}
\author{G.\ Goll}
\email{gernot.goll@kit.edu}
\affiliation{Physikalisches Institut, Karlsruher Institut f\"ur Technologie, D-76131
Karlsruhe, Germany}
\begin{abstract}
We report on investigations of the superconducting energy gap of the $A15$ superconductor Nb$_3$Sn by point-contact spectroscopy of breakjunctions. The voltage-dependent differential conductance $dI/dV$ reveals features of a second energy gap besides the energy gap known from previous tunnel measurements with maxima at $\Delta_1=3.92\pm 0.16\,$meV and $\Delta_2=0.85\pm 0.17\,$meV as derived from a histogram summarizing the data of more than 60 contacts. These findings are the first spectroscopic evidence that Nb$_3$Sn belongs to the class of two-band superconductors and they are in line with low-temperature specific-heat measurements on Nb$_3$Sn. 
\end{abstract}  
\date{\today}
\maketitle
Multi-band superconductivity is an outstanding issue of recent research. Although theoretical considerations last back already to early 1960s~\cite{suhl59,schopohl77}, for a long time there was no experimental evidence of its realization in nature. The discovery of two-band superconductivity in MgB$_2$ has revitalized the interest in multigap superconductivity. A first experimental hint at the opening of multiple gaps at the Fermi energy is a low-temperature anomaly in the specific heat $c(T)$ in the superconducting state of the superconductor. Instead of the predicted exponential decay below $T_c$ according to single-gap BCS theory~\cite{bcs57}, a plateau at low temperatures occurs. In MgB$_2$ the specific heat can be satisfactorily explained in a two-band superconductor model~\cite{bouquet01}, which considers the contribution of each band independently for negligible interband scattering. 
Recently, measurements of the specific heat of Nb$_3$Sn have shown a similar behavior, namely the non-exponential decay and a plateau at low temperatures. It has been quantitatively ascribed to the presence of a second superconducting gap~\cite{guri04}, which affects only $7.5\,\%$ of the electronic density of states. We performed point-contact spectroscopy on Nb$_3$Sn in order to directly probe the structure of the superconducting energy gap spectroscopically. This method is a useful tool to clearly identify different energy gaps as it was done, e.\,g. for MgB$_2$~\cite{laub01,yanson04}, and to investigate gap anisotropies~\cite{goll06a}. The sensitivity of point-contact spectroscopy on the energy gap results from almost ballistic quasiparticles which probe the local density-of-states at the superconductor-normal metal or superconductor-superconductor interface of a constriction. Our measurement confirm the existence of a second energy gap in Nb$_3$Sn.\\
Nb$_3$Sn is a strong coupling $s$-wave superconductor with a critical temperature $T_c\approx18\,$K and an upper critical field $H_{c2}\approx 25 \, \rm{T}$. It is a technically relevant superconductor used, e.g.\, in magnet coils. A recent review summarizes the properties of Nb$_3$Sn from a metallurgical point of view~\cite{godeke06}.
\begin{figure}[hbt]
\begin{center}
\includegraphics[width=\picwidth]{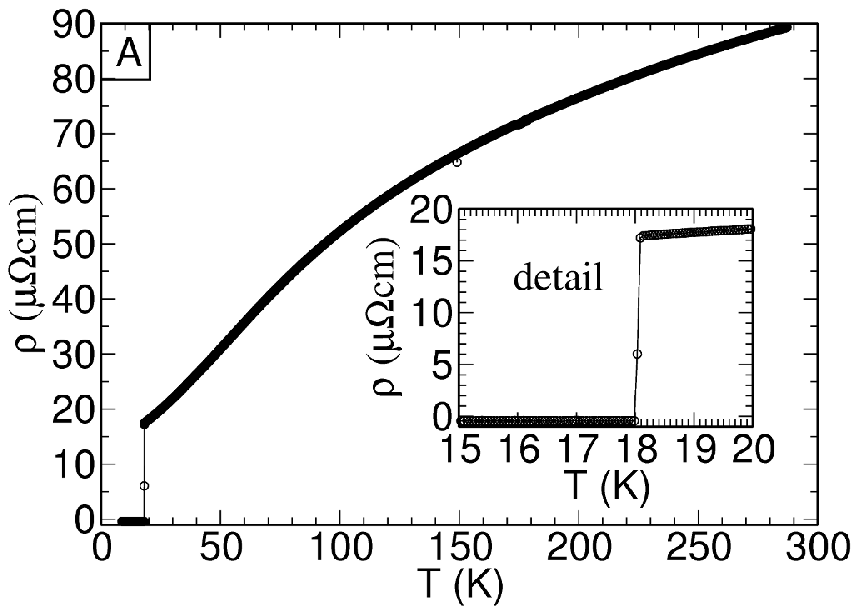}\\
\vspace{5mm}
\includegraphics[width=\picwidth]{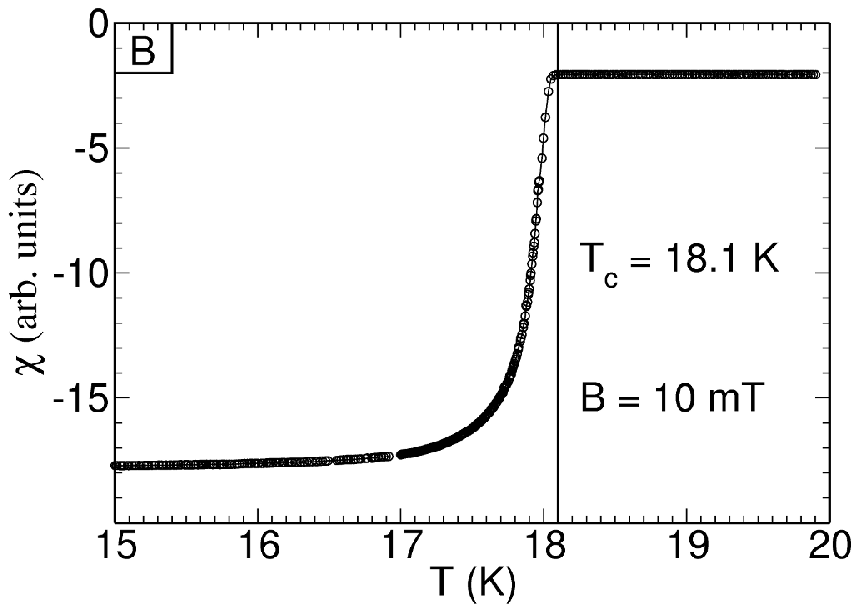} 
\end{center}
\caption{(A) Measurement of the resistivity vs.\ temperature from room temperature to $T=1.5\,$K. The inset shows the superconducting transition at $T_c=18.1\,$ in detail. The transition is very sharp with $\Delta T^{90-10\%}_c =0.15\,$K.\label{fig:rho}\\ 
(B) Measurement of the susceptibility vs.\ temperature, zero-field-cooled-field-heated curve for $B=10\,$mT. The onset of the transition to superconductivity is determined to be $T_c=18.1\,\rm{K}$ and is in good agreement with the resistivity measurement.\label{fig:vsm}}
\end{figure}
Early measurements of the electronic structure of Nb$_3$Sn were done by the use of tunnel spectroscopy on thin-film tunnel junctions~\cite{shen72,moore79,wolf80,rudman84,geerk85,geerk86}. A satisfactory interpretation of the tunnel data always had to postulate a proximity-induced surface layer or normal conducting grains in order to explain the obvious features at low bias in terms of leakage currents. We used a polycrystalline bulk sample synthesized at the Karlsruhe Institute of Technology (Campus North) in a powder metallurgical approach under hot isostatic pressing (HIP). The powder metal approach results in very homogeneous samples with very sharp transitions at relatively high $T_c$ values, for details see ref.~\onlinecite{gold93}. We verified the high quality of the sample by measurements of the resistivity $\rho$ and the susceptibility $\chi$ as function of temperature. For the resistivity and the point-contact measurements small bars with a size of about $6.0 \times 1.2\times 0.15\,$mm$^3$ were cut by spark erosion from a larger polycrystal.
The resistivity measurements have been performed in a four-point geometry within a temperature range from room temperature (RT) down to $T=1.5\,$K. The result is depicted in figure~\ref{fig:rho}(A). The inset shows an enlargement in the vicinity of the superconducting transition. The very sharp step with $\Delta T^{90-10\%}_c =0.15\,$K at the transition temperature $T_c=18.1\,$K is an evidence for the high quality of the sample. The resistivity ratio RR$=\rho(RT)/\rho(T_c)=5.2$ and $\rho(T_c)=17.2\,\mu\Omega\mathrm{cm}$, the residual resistivity at $T_c$, are in reasonable agreement with previously reported values~\cite{gold93}, however, no feature caused by the cubic-tetragonal transition is seen in our resistivity data. Figure~\ref{fig:vsm}(B) shows $\chi (T)$ data obtained with a commercially available vibrating sample magnetometer (VSM). The curve shows the zero field cooled - field heated measurement in an applied field of $B=10\,$mT. The onset of the transition can be identified to be at $T_c=18.1\,$K. The long tail at lower temperatures is due to the fact that Nb$_3$Sn is a type-II superconductor with a small lower critical field.\\
First point-contact measurements were performed on superconductor-normal metal contacts with Pt as the normal conducting counter electrode~\cite{goll06b}. These data already show the existence of two gap-related features at $\Delta_1=2.0\,$meV and $\Delta_2=0.4\,$meV in the $dI/dV$ vs. $V$ spectra. However, the size of the major energy gap is considerable lower than the value $\Delta=3.15\,$meV reported in literature~\cite{geerk86}. The reduced gap value is a consequence of the degradation of the superconducting properties at the surface caused by oxidation of the sample surface. To avoid such surface effects in the contact region we used the breakjunction technique to prepare Nb$_3$Sn point contacts with a fresh, non corrupted interface. On a tiny bar of Nb$_3$Sn with a notch to predetermine the breaking point we attached four leads. The assembly was glued on an insulating substrate by use of stycast epoxy resin and then mounted to the low-temperature end of our point-contact device. In our measuring system we were able to break the bar $in$ $situ$ at low temperature and under a $^4$He atmosphere, resulting in Nb$_3$Sn-Nb$_3$Sn homo contacts, i.\,e.\ superconductor/superconductor point contacts with a fresh interface. The breakjunction resistance is varied by opening and closing the junction in a controlled fashion at low temperatures and the contact resistance is monitored during the whole measurement series. We have measured the differential conductance of more than $60$ contacts on three different bars cut from the same sample. The experimental setup allows us to reach temperatures down to $T=1.5\,\rm{K}$ and magnetic fields up to $B=6\,\rm{T}$. Measurements of the current-voltage characteristics as well as spectra of the differential conductance $dI/dV$ vs.\ voltage $V$ were recorded simultaneously in a standard four-point geometry of the leads. To measure the $dI/dV$-characteristics lock-in technique was used. For that purpose, the current applied to the sample was modulated by a small ac current with $\nu=9.13\,$kHz.\\
\begin{figure}[hbt]
\includegraphics[width=\picwidth]{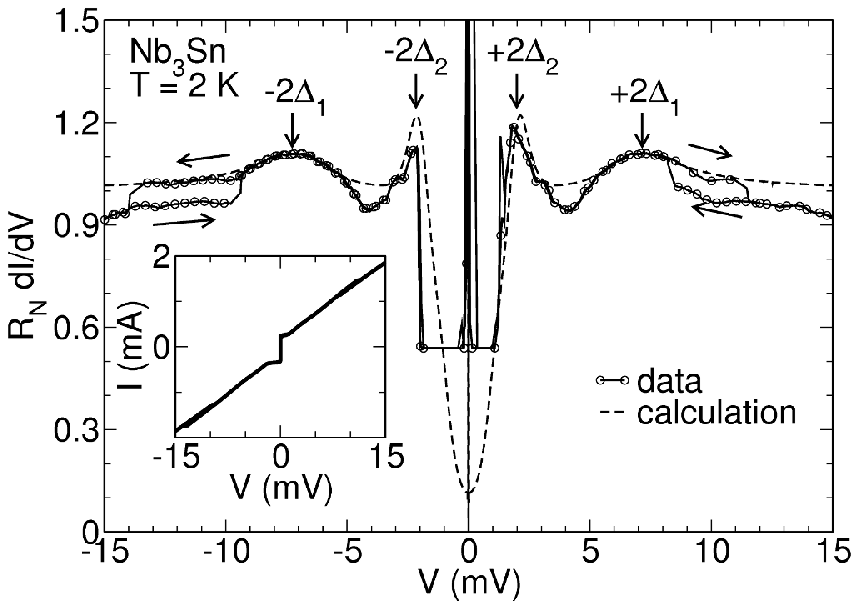}
\caption{Normalized differential conductance $R_N \cdot dI/dV$ vs.\ voltage $V$. The circles show the measurement, the features ascribed to two-band superconductivity are marked by vertical arrows. The large anomaly at zero bias is due to the Josephson effect. The hysteretic conductivity cange for $|V|=8-13\,$mV is probably caused by a rearrangement of the contact in the electric field. The sweeping direction is marked by horizontal arrows. The dashed curve is calculated within a simple model for two-band superconductivity, neglecting interband scattering and the Josephson effect. The inset shows the current vs.\ voltage spectra, showing the typical jump in the current for low bias due to the Josephson effect.\label{fig:data52}}
\end{figure}
For one-band superconductor-superconductor homo contacts one expects in a simple semiconductor model a single characteristic peak in the differential conductance vs.\ voltage spectra at the voltage value $V= 2\Delta/e$ for each polarity, where $\Delta$ is the energy gap given by $\Delta=1.75\,$k$_{\mathrm{B}}$T$_c$ in the BCS weak-coupling limit. In contrast to this, in most of our measurements we could find at least two of those pronounced characteristic peaks for each polarity although the feature can be more or less distinctive when varying the contact resistance. Figure~\ref{fig:data52} shows a representative spectrum for a contact with $R_N=12.5\,\mathrm{\Omega}$ measured at $T=2\,$K (circles). The inset shows the corresponding current-voltage characteristic. The dashed curve in figure~~\ref{fig:data52} was calculated according to a standard two-gap approach (see below). Some features of minor importance are the hysteresis at $|V|=8-13\,$mV and the large anomaly at zero bias. The hysteretic feature was observed for a few contacts only. It indicates a tiny chance of the contact which is caused probably by a rearrangement of atoms or cluster of atoms in the break-junction region in the applied electric field.  The large anomaly at zero bias is observed for about $15\,$\% of all junctions. It is caused by Josephson coupling of the two superconducting electrodes. The Josephson current is clearly identifiable in the current-voltage spectra (see inset of Fig.~\ref{fig:data52}).\\
For a theoretical description of the quasiparticle current through the junction the currents attributed to each band, respectively superconducting gap are, considered to be independent from each other. If interband scattering is neglected, the total current $I_{\textrm{tot}}(V)$ through the point-contact can be written as the sum of the two contributions:
\begin{equation*}
 I_{\textrm{tot}}(V)=\alpha \cdot I_{\Delta_1}^{SS}(V)+(1-\alpha)\cdot I_{\Delta_2}^{SS}(V)
\end{equation*}
with a weighting parameter $\alpha$ ($0< \alpha <1$). Each contribution $I_{\Delta_i}^{SS}(V)$ was calculated according to 
\begin{eqnarray*}
 I_{\Delta_i}^{SS}(V)&=\\ 
 \frac{G_{nn}}{e}\int\limits_{-\infty}^{+\infty}\frac{
N^{i}_{s}(E)N^{i}_s(E+eV)}{N(0)^2}&\left(f(E)-f(E+eV)\right)
\end{eqnarray*}
where $N^{i}_{s}(E)/N(0)=E/\sqrt{E^2-\Delta_i^2}$ for $E>\Delta_i$ and $N^{i}_{s}(E)/N(0)=0$ for $E<\Delta_i$. In addition, a Dynes parameter $\Gamma$ was included to account for inelastic scattering processes in the contact region, i.e.\ $E\rightarrow E+i\Gamma$. The total normalized differential conductance 
\begin{eqnarray*}
 \left[dI/dV\right]_{\textrm{tot}}(V)&= \\
\alpha \left[dI/dV\right]_{\Delta_1}(V)&+(1-\alpha) \left[dI/dV\right]_{\Delta_2}(V)
\end{eqnarray*}
was calculated by numerically differentiation of the total current $I_{\textrm{tot}}(V)$ with respect to $V$. The dashed line in figure~\ref{fig:data52} was calculated in this manner. Of course that description neglects the presence of the Josephson feature in the spectra, c.f. in Fig.~\ref{fig:data52}. The parameters that describe best the data in figure~\ref{fig:data52} are:
$$ \Delta_1=3.4 \,\rm{meV} \quad \rm{and} \quad \Delta_2=1.0 \,\rm{meV} $$
with a weighting parameter $\alpha=0.22$. Inelastic scattering processes in the contact were parameterized by $\Gamma_i=0.32\,\Delta_i$.\\
\begin{figure}[hbt]
\includegraphics[width=\picwidth]{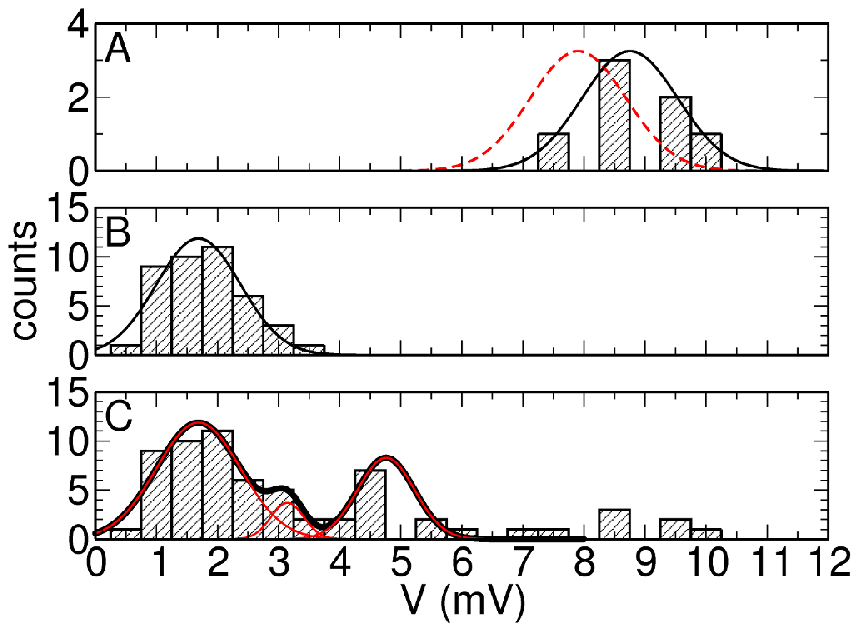}
\caption{Histograms showing the occurrence of the peak voltage values, the solid and dashed lines are Gaussian shaped fits to the data. The maxima of these Gaussian shaped curves represent the expected voltage values where features of the energy gap in the spectra are expected. Structures we ascribed to the large or small gap are show separate in A and B, respectively. Histogram C summarizes all measured structures.\label{fig:histo}}
\end{figure}
The measurements of about twenty contacts for $T<2\,\rm{K}$ were analyzed and the voltage values at which the peaks in $dI/dV$ appeared were summarized in three histograms, see figure~\ref{fig:histo}. In these histograms we can find different accumulation points. The accumulation points are highlighted by Gaussian shaped curves that are fitted to the data. Applying the simple semiconductor model to the case of two-band superconductivity one expect features in the differential conductance of the superconductor-superconductor contact at $V_1=2\Delta_1/e$ and $V_2=2\Delta_2/e$ as well as for $V_3=(\Delta_1+\Delta_2)/e$ and $V_4=(\Delta_1-\Delta_2)/e$. Therefore, panel A displays the data we associated with the large gap at $V_1$, panel B displays the data we associated with the small gap at $V_2$. We can determine the values of the energy gap ascribed to each of the accumulation points found in the histogram from the Gaussian fits. This results in:
\begin{equation*}
\Delta_1= 4.38\pm 0.03\,\rm{meV} \quad \rm{and} \quad \Delta_2=0.86\pm 0.05\,\rm{meV}
\end{equation*}
when fitting the data in figure~\ref{fig:histo} A and B (solid lines). As a cross-check the energy gaps can also be obtained by fitting three Gaussian lines to the data between $0$ and $6.5\,$mV in figure~\ref{fig:histo} C, i.e.\ to the peaks associated with $V_2$, $V_3$, $V_4$. This results in 
\begin{equation*}
\Delta_1= 3.92\pm 0.16\,\rm{meV} \quad \rm{and} \quad \Delta_2=0.85\pm 0.17\,\rm{meV},
\end{equation*}
i.e.\ a comparable $\Delta_2$ and a little lower $\Delta_1$.
For comparison the dashed line in figure~\ref{fig:histo} A shows a Gaussian calculated with the gap value $2\Delta_1/e$ obtained from the three Gaussians fits to Fig.~\ref{fig:histo}c.
$\Delta_1$ is is somewhat larger than that obtained in previous tunneling measurements on thin-film tunnel junctions (see table~\ref{table}), but $\Delta_2$ is the first spectroscopic confirmation of the existence of a second gap in Nb$_3$Sn. Both values are also a little larger than those derived from the specific-heat measurement, where $\Delta_1 = 3.68 \,$meV and $\Delta_2 = 0.61 \,$meV have been reported~\cite{guri04}. However, we note that determining $\Delta$ from the peak in $dI/dV$ at finite temperature overestimates the gap value. Only at $T=0$ the maxima appear at $\pm 2\Delta/e$. The identification of a second energy gap also sheds new light on the interpretation of the low-voltage feature in previous tunneling data.\\
\begin{table*}
\caption{\label{table}}
\begin{tabular}[c]{p{0.4\linewidth} p{0.15\linewidth} p{0.15\linewidth} p{0.15\linewidth} p{0.1\linewidth} }
\hline Experiment & $T_c\,$(K)  & $\Delta_1\,$(meV) & $\Delta_2\,$(meV) & Ref. \\ 
\hline 
 Nb$_3$Sn / oxidized a-Si / Pb tunnel junction & 17.7 & 3.4 & - &  \onlinecite{moore79} \\  
 Nb$_3$Sn / oxidized a-Si / Pb tunnel junction & 17.4 & 3.30 & - &  \onlinecite{rudman84}\\ 
 Nb$_3$Sn / AlZr-oxid / Pb tunnel junction & 18 & 3.13 & - &   \onlinecite{geerk85, geerk86}\\ 
 Specific heat & 18 & 3.68 & 0.61 &   \onlinecite{guri04}\\ 
 Nb$_3$Sn / Pt point contact & 18.1 & 2.0 & 0.4 &   \onlinecite{goll06b}\\ 
 Nb$_3$Sn - Nb$_3$Sn breakjunction & 18.1 & 3.92 & 0.85 & this work \\ 

\hline
\end{tabular}
\end{table*}
We also studied the temperature dependence of the characteristic peaks in the $dI/dV$ spectra. The structure that we assign to be caused by the large gap can be followed over the whole temperature range up to $T_c$. For increasing temperature, the height of the peaks is reduced, the peaks become broader and shift to lower voltage values as expected. The peaks that we associate with the small energy gap cannot be clearly resolved for temperatures $T\ge8\,$K. We attribute this to the huge zero bias feature caused by the supercurrent and to thermal broadening of the structures, both interfering with the peaks of the small energy gap.
 \begin{figure}[hbt]
 \includegraphics[width=\picwidth]{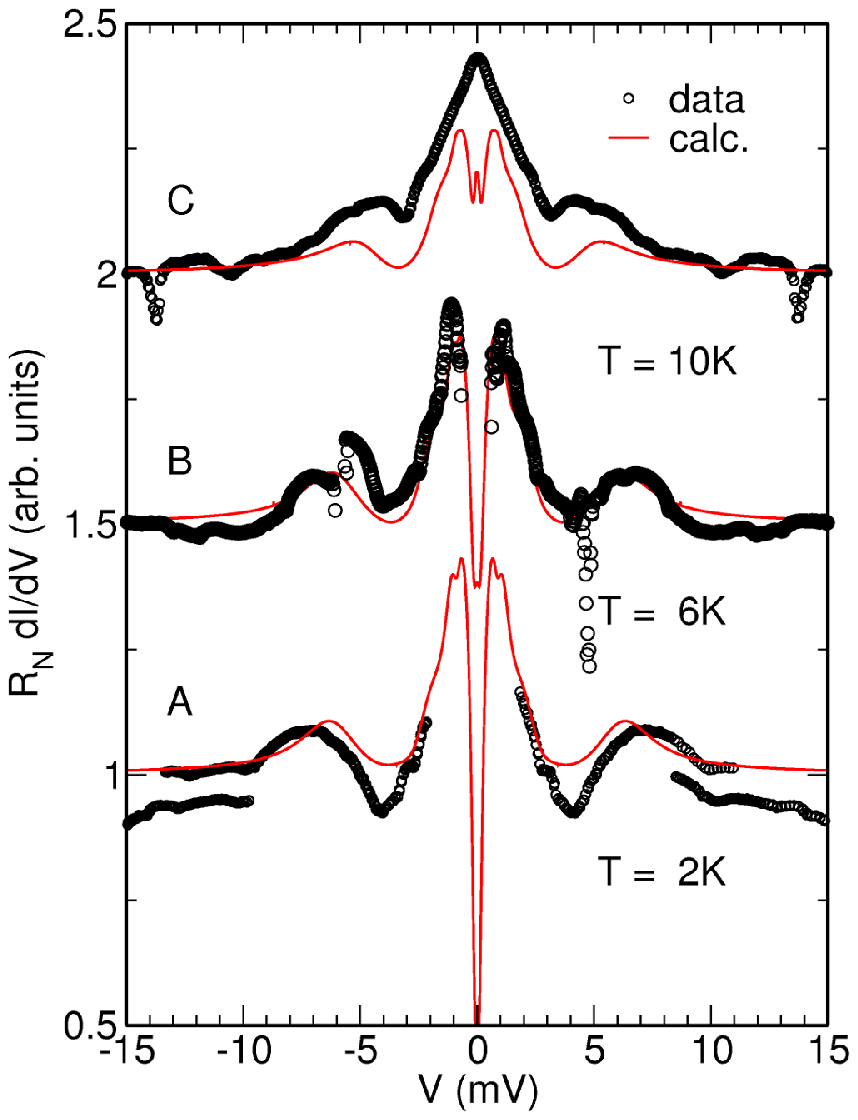}
 \caption{Temperature dependence of the S-S-contact shown in figure~\ref{fig:data52}, measurements (circles) for $T=2\,$K (A), $T=6\,$K (B) and $T=10\,$K (C) are shown as examples. The fit (solid line) was done for the $T=6\,$K measurement, the parameters were kept constant in the calculation for the other temperatures. For the calculations, the temperature dependency of the gap was assumed to follow BCS behavior. For clarity, the curves are shifted against each other.}
 \label{fig:tsweep}
 \end{figure}
The results for three different temperatures are shown in figure~\ref{fig:tsweep}, the data are marked with circles and are shifted against each other for clarity. The solid lines are calculated by the two-gap approach as described above. The data obtained for $T=6\,$K were fitted to determine the parameters $\Delta_1$, $\Delta_2$, $\alpha$ and $\Gamma$. For the other temperatures these parameters were kept constant and a BCS dependency of the energy gap has been assumed. As one can see in figure~\ref{fig:tsweep}, the temperature dependence is not exactly what is expected according to BCS theory. The deviations are more significant for the higher temperatures. Nevertheless the tendency of the peaks to shift to lower voltage values can be clearly seen. For temperatures $T\ge T_c$ the gap features vanish and the differential conductance gets constant as expected.\\
We also investigated the influence of an applied magnetic field on both gap features. In our experimental setup we had the possibility to achieve magnetic fields up to $H=6\,$T, which is only $0.25 \, H_{c2}$. Therefore, only a small influence of the magnetic field on the energy gap is expected which can hardly be distinguished in the experimental spectra. However, the Josephson current is completely suppressed in magnetic field.\\ 
In order to exclude non-intrinsic origins for the observation of a second energy gap, we finally checked the sample for possible superconducting impurity phases Nb$_6$Sn$_5$ ($T_c=2.07$\,K), NbSn$_2$ ($T_c=2.68$\,K) \cite{cha66} and for elemental Sn ($T_c=3.72$\,K) with expected gap values $\Delta\approx 0.3-0.6\,$meV close to the experimental value of the small energy gap. X-ray structure analysis made on an equivalent sample piece shows tiny peaks of an unknown phase \cite{rem1}. The peaks cannot be allocated to known structure data of above impurity phases and the phase comprises 1-2\% of the volume at most. To ensure, that there are no impurity phases at the breakjunctions interface, we analyzed the surface of the contact region by meaning of scanning electron microscopy (SEM) and used energy-dispersive X-ray (EDX) to determine the elements at the surface. No elemental Sn (and Nb) inclusions were located in the EDX mapping. A uniform distribution of Nb and Sn has been found and the stoichiometric content of Nb$_3$Sn could be confirmed within the accuracy of the EDX. Therefore, we can exclude that elemental Sn causes the gap features assigned to the small gap. In addition, Sn has a very low critical field and in an applied field one expects that the features caused by the presence of the small gap would be immediately suppressed. This is not observed in the experiment.\\
In conclusion, our point-contact measurements on break junctions of Nb$_3$Sn for the first time show the existence of two superconducting energy gaps in this material spectroscopically. The size of both gaps is in quantitative agreement with the specific-heat measurements previously reported. The identification of Nb$_3$Sn as a two-band superconductor with two independent methods asks for a revisiting of the physical properties of the A15 compounds. We thank G. Pfundstein for the SEM and EDX measurements and A. Jung for additional EDX and X-ray analysis on a second equivalent sample. We acknowledge helpful discussions with J. Geerk and H. v. L\"ohneysen.

\end{document}